\documentclass[%
 superscriptaddress,
preprint,
 amsmath,amssymb,
 aps,
]{revtex4-1}

\usepackage{graphicx}
\usepackage{dcolumn}
\usepackage{bm}
\usepackage{color}
\usepackage{hyperref}
\usepackage{siunitx}
\usepackage{xfrac}

\begin{document}

\title{Sub-nanotesla magnetometry with a fibre-coupled diamond sensor}

\author{R. L. Patel}
\email{Rajesh.Patel@warwick.ac.uk}
\affiliation{Department of Physics, University of Warwick, Gibbet Hill Road, Coventry CV4 7AL, United Kingdom}
\affiliation{EPSRC Centre for Doctoral Training in Diamond Science and Technology, University of Warwick, Coventry, CV4 7AL, United Kingdom}
\author{L. Q. Zhou}
\affiliation{Department of Physics, University of Warwick, Gibbet Hill Road, Coventry CV4 7AL, United Kingdom}
\author{A. C. Frangeskou}
\thanks{Current address: Lightbox Jewelry, Orion House, 5 Upper St. Martin’s Lane, London, WC2H 9EA, United Kingdom}
\affiliation{Department of Physics, University of Warwick, Gibbet Hill Road, Coventry CV4 7AL, United Kingdom}
\author{G. A. Stimpson}
\affiliation{Department of Physics, University of Warwick, Gibbet Hill Road, Coventry CV4 7AL, United Kingdom}
\affiliation{EPSRC Centre for Doctoral Training in Diamond Science and Technology, University of Warwick, Coventry, CV4 7AL, United Kingdom}
\author{B. G. Breeze}
\affiliation{Department of Physics, University of Warwick, Gibbet Hill Road, Coventry CV4 7AL, United Kingdom}
\affiliation{Spectroscopy RTP, University of Warwick, Gibbet Hill Road, Coventry CV4 7AL, United Kingdom}
\author{A. Nikitin}
\thanks{Current address: School of Engineering, University of Warwick, Coventry, CV4 7AL, United Kingdom}
\affiliation{Department of Physics, University of Warwick, Gibbet Hill Road, Coventry CV4 7AL, United Kingdom}
\author{M. W. Dale}
\thanks{Current address: De Beers Group Technology, Belmont Rd, Maidenhead SL6 6JW, United Kingdom}
\affiliation{Department of Physics, University of Warwick, Gibbet Hill Road, Coventry CV4 7AL, United Kingdom}
\author{E. C. Nichols}
\affiliation{Department of Physics, University of Warwick, Gibbet Hill Road, Coventry CV4 7AL, United Kingdom}
\author{W. Thornley}
\affiliation{Department of Physics, University of Warwick, Gibbet Hill Road, Coventry CV4 7AL, United Kingdom}
\author{B. L. Green}
\affiliation{Department of Physics, University of Warwick, Gibbet Hill Road, Coventry CV4 7AL, United Kingdom}
\author{M. E. Newton}
\affiliation{Department of Physics, University of Warwick, Gibbet Hill Road, Coventry CV4 7AL, United Kingdom}
\affiliation{EPSRC Centre for Doctoral Training in Diamond Science and Technology, University of Warwick, Coventry, CV4 7AL, United Kingdom}
\author{A. M. Edmonds}
\affiliation{Element Six Innovation, Fermi Avenue, Harwell Oxford, Didcot, OX11 0QR, Oxfordshire, United Kingdom}
\author{M. L. Markham}
\affiliation{Element Six Innovation, Fermi Avenue, Harwell Oxford, Didcot, OX11 0QR, Oxfordshire, United Kingdom}
\author{D. J. Twitchen}
\affiliation{Element Six Innovation, Fermi Avenue, Harwell Oxford, Didcot, OX11 0QR, Oxfordshire, United Kingdom}
\author{G. W. Morley}
\email{Gavin.Morley@warwick.ac.uk}
\affiliation{Department of Physics, University of Warwick, Gibbet Hill Road, Coventry CV4 7AL, United Kingdom}
\affiliation{EPSRC Centre for Doctoral Training in Diamond Science and Technology, University of Warwick, Coventry, CV4 7AL, United Kingdom}

\begin{abstract}
Sensing small magnetic fields is relevant for many applications ranging from geology to medical diagnosis. We present a fiber-coupled diamond magnetometer with a sensitivity of (310~$\pm$~20)~pT$/\sqrt{\text{Hz}}$ in the frequency range of 10-150~Hz. This is based on optically detected magnetic resonance of an ensemble of nitrogen vacancy centers in diamond at room temperature. Fiber coupling means the sensor can be conveniently brought within 2~mm of the object under study.

\end{abstract}

\date{\today}
\maketitle

\section{Introduction}

\begin{figure} 
\includegraphics[width=1.0\textwidth]{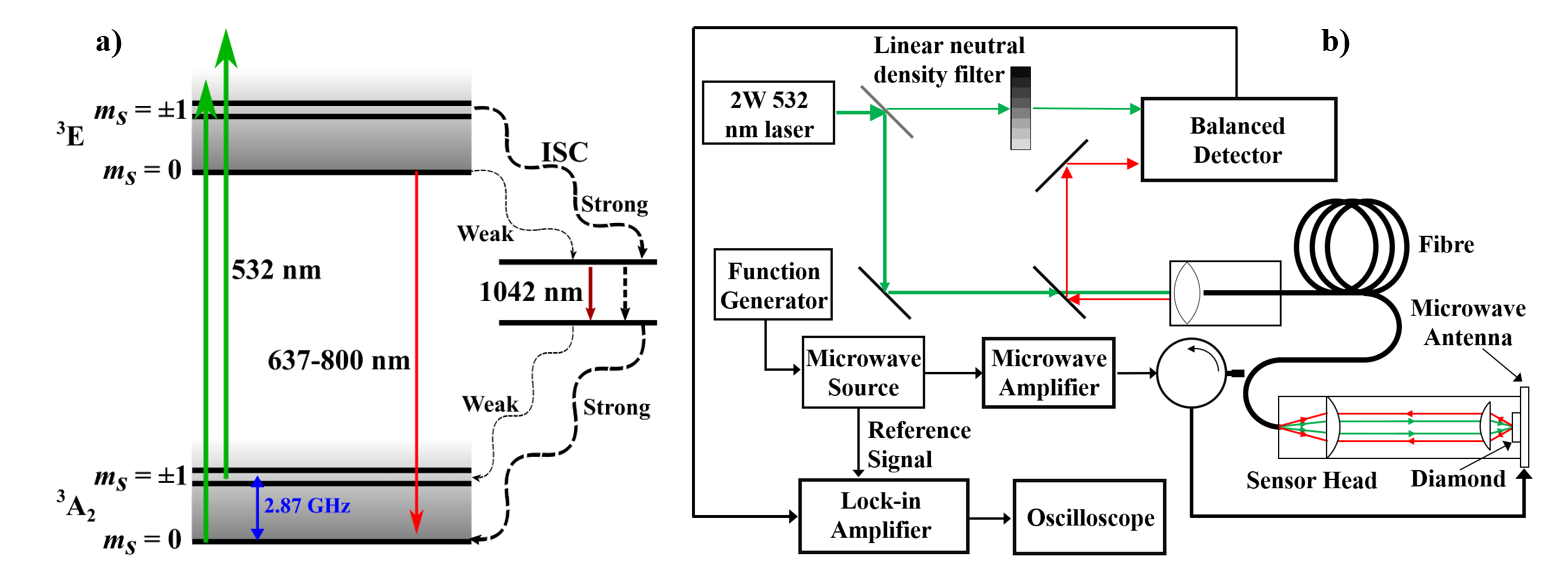} 
\caption{a) Energy level diagram of the nitrogen-vacancy center (NVC) in diamond, radiative and non-radiative transitions are indicated by solid lines and dashed lines respectively. A 532~nm laser excites the system from the $^{3}$A$_{2}$ states to the $^{3}$E states. The $m_{s}$~=~$\pm$1 states are more likely than the $m_{s}$~=~0 state to decay via the inter-system crossing (ISC) leading to spin polarisation into the $m_{s}$~=~0 state. The $^{3}$E state can decay back to the $^{3}$A$_{2}$ by emitting 637~nm to 800~nm light and the intensity is greater for the $m_{s}$~=~0 state allowing for optical detection of magnetic resonance. b) A schematic of our fiber-coupled magnetometer.}
\label{NV_diagram}
\end{figure}

The sensing of magnetic fields using the nitrogen vacancy center (NVC) in diamond has seen rapid growth over the last decade due to the promise of high sensitivity magnetometry with exceptional spatial resolution~\cite{Balasubramanian2008,Taylor2008a} along with a high dynamic range~\cite{Clevenson2018}. The use of NVC ensembles rather than single centres improves sensitivity while degrading the spatial resolution~\cite{Pham_2011,Barry14133,Schloss2018,PhysRevB.85.121202,Levine2019,Barry2019,Eisenach2018,Bougas2018,Clevenson2015,Clevenson2018,Kim2019}. Recent advancements have demonstrated ensemble sensitivities of 0.9~pT$/\sqrt{\text{Hz}}$ for d.c. fields~\cite{Fescenko2019} and 0.9~pT$/\sqrt{\text{Hz}}$ for a.c fields~\cite{Wolf2015}. However, these results have been limited to systems that are bulky and are typically fixed to optical tables. In contrast, fibre-coupling provides a small sensor head that may be moved independently from the rest of the control instrumentation and thus offers the possibility of application in medical diagnostic techniques such as magnetocardiography (MCG)~\cite{Dale2017,Fenici2005}. Most fiber-coupled diamond magnetometers have relied on using nanodiamonds/microdiamonds attached to the end of a fiber, achieving sensitivities in the range of 56000-180~nT$/\sqrt{\text{Hz}}$~\cite{Liu2013,Fedotov2014a,Duan:19}. Utilising a two-wire microwave transmission line in addition to a fiber-diamond set-up was able to achieve a sensitivity of $\sim$300~nT$/\sqrt{\text{Hz}}$~\cite{Fedotov2014}. A fiber-based gradiometer approach was able to provide a sensitivity of $\sim$35~nT$/\sqrt{\text{Hz}}$ with projected shot-noise sensitivities potentially allowing for MCG~\cite{Blakley2016,Blakley2015}. Using a hollow-core fiber with many nanodiamond sensors in a fluidic environment provided a sensitivity of 63~nT$/\sqrt{\text{Hz}}$ per sensor and a spatial resolution of 17~cm~\cite{doi:10.1002/lpor.201900075}. Other compact magnetometers that use a fiber have demonstrated sensitivities in the ranges of 67-1.5~nT$/\sqrt{\text{Hz}}$~\cite{10.1088/1361-6463/ab6af2,Webb2019,Dmitriev:16}. The best sensitivity reported for a fibre-coupled diamond magnetometer so far is 35~nT$/\sqrt{\text{Hz}}$ when sensing a real test field~\cite{Blakley2016}, and 1.5~nT$/\sqrt{\text{Hz}}$ when estimating the sensitivity based on the signal-to-noise-to-linewidth using the slope of a resonance in the magnetic resonance spectrum~\cite{Dmitriev:16}. Other diamond magnetometers which offer high portability whilst maintaining a compact structure have been demonstrated with a compact LED-based design achieving a minimum detectable field of $\SI{1}{~\micro\tesla}$ whilst offering minimal power consumption ~\cite{STURNER201959}. Here, a diamond-based fiber-coupled magnetometer with sub-nT sensitivity is presented. The key feature is the use of lenses to reduce optical losses from the fiber to the diamond and back as shown in figure~\ref{NV_diagram}b).

The NVC, when in its negative charge state, is a spin $S = 1$ defect that can be optically initialised into the $m_s = 0$ ground state and possesses spin-dependent fluorescence giving rise to optically detected magnetic resonance (ODMR) \cite{Doherty2013}. The energy level diagram is shown in figure~\ref{NV_diagram}a). The Zeeman-induced splitting of the NVC leads to the detection of magnetic fields with high sensitivity where the sensitivity of the magnetometer scales with $1/ \sqrt{N}$, where $N$ is the number of centers probed~\cite{PhysRevB.80.115202,Rondin2014}. The zero-field splitting at room temperature is $\sim$2.87~GHz. Upon application of an external magnetic field Zeeman-induced splitting leads to sub-levels that are split by

\begin{equation}
    \Delta f = \frac{2 g_{e} \mu_{B} B_{||}}{h}
\end{equation}

\noindent where $g_e = 2.0028$ is the NVC g-factor, $\mu_B$ is the Bohr magneton, $B_{||}$ is the projection of the external magnetic field onto the NVC symmetry axis (the $\langle 111 \rangle$ crystallographic direction) and $h$ is Planck's constant. The energy levels are further split by the hyperfine interaction between the electron spin and $^{14}$N nuclear spin ($I = 1$) by $A \approx 2.16$~MHz. Under a continuous wave excitation scheme, which is employed in this paper, the photon-shot-noise-limited sensitivity of a diamond-based magnetometer is given by

\begin{equation} \label{PSNL}
    \eta = \frac{4}{3\sqrt{3}} \frac{h}{g_{e} \mu_{B}} \frac{\Delta \nu}{C \sqrt{I_{0}}},
\end{equation}

\noindent where $\Delta \nu$ is the linewdith, $C$ is the measurement contrast (the reduction in fluorescence when on resonance compared to when not on resonance) and $I_{0}$ is the number of collected photons off resonance~\cite{Dreau2011,Barry14133}.

\section{Methods}

Magnetometry is performed with the set-up shown in figure~\ref{NV_diagram}b). A Laser Quantum Gem-532 with a maximum power output of 2~W is used to excite the NVC ensemble; for our experiments 1~W was used to reduce laser noise. The laser beam is passed through a Thorlabs BSF10-A beam sampler whereby approximately 1\% is picked off and supplied to the reference arm of a Thorlabs PDB450A balanced detector to cancel out laser intensity noise; the illumination levels incident upon each photodiode is equal in the absence of microwaves. The remaining (high-intensity) portion of the laser beam is focused into a custom-ordered 5~m 0.22 N.A. Thorlabs FG400AEA fiber with a core diameter of 400~\si{\micro\meter} and ceramic FC/PC termination. The fiber output is focused onto the diamond using a pair of aspheric lenses (Thorlabs) housed in a adjustable SM1 tube lens (Thorlabs SM1NR05). The first lens (C171TMD-B) collimates the fiber output whilst the second (C330TMD-B) focuses the beam onto the diamond. The same lenses are used to collect the emitted fluorescence from the NVC ensemble.

Microwave excitation is provided by an Agilent N5172B microwave source with a Mini-Circuits ZHL-16W-43-S+ microwave amplifier; hyperfine excitation is utilized to improve the contrast by mixing a 2.158~MHz sinewave~\cite{Barry14133} from a RSPro arbitrary function generator AFG21005. The microwaves are then delivered to a 2-4 GHz coaxial circulator to reduce microwave reflection. The microwaves are square-wave frequency modulated~\cite{El-Ella2017}. The signal from the balanced detector is supplied to a Zurich MFLI DC - 500 kHz lock-in amplifier (LIA). The diamond is mounted on a 5~mm copper loop deposited onto an aluminium prototyping board (C.I.F AAT10) for microwave delivery and heat management. A permanent rare earth magnet is aligned to the (111) crystallographic orientation of the NVC ensemble, proving the spectrum shown in figure~\ref{LIA_ODMR}a). The optimum sensitivity was determined by varying the microwave power, the modulation amplitude and the frequency modulation to find the maximum value of the zero-crossing slope, as shown in figure~\ref{LIA_ODMR}. The microwave power was varied from -25 dBm to -8 dBm (pre-amplification value before the 43~dB amplifier, corresponding to 0.06~W and 3.16~W respectively after amplification with losses due to the cables neglected) in 1 dBm increments. The frequency modulation was investigated between 1 to 80~kHz and the frequency deviation was varied from 100 to 600~kHz. All ODMR spectra are taken with a frequency sweep range of 2.76~GHz to 2.94 GHz, a step resolution of 20~kHz, a step dwell time of 4~ms and a LIA frequency bandwidth of 200~Hz using a 48~dB/octave filter slope; the LIA output scaling was set to 500.

The microwave frequency is set then to the zero-crossing point and the fluorescence is monitored with the sensitivity determined by considering the mean and standard deviation of 160 one-second fast-Fourier transforms (FFT's) using an oscilloscope. The sample employed in this work is a single crystal 99.995\% $^{12}$C enriched sample of dimension 4~mm~$\times$~4~mm~$\times$~0.6~mm with a (100)-orientation, grown through chemical vapour deposition by Element Six. The concentration of negatively charged NVC, see appendix~\ref{Diamond_characterisation}, was measured to be [NV$^{-}$]~=~4.6~ppm. 

\section{Results}

\begin{figure}
\includegraphics[width=\textwidth]{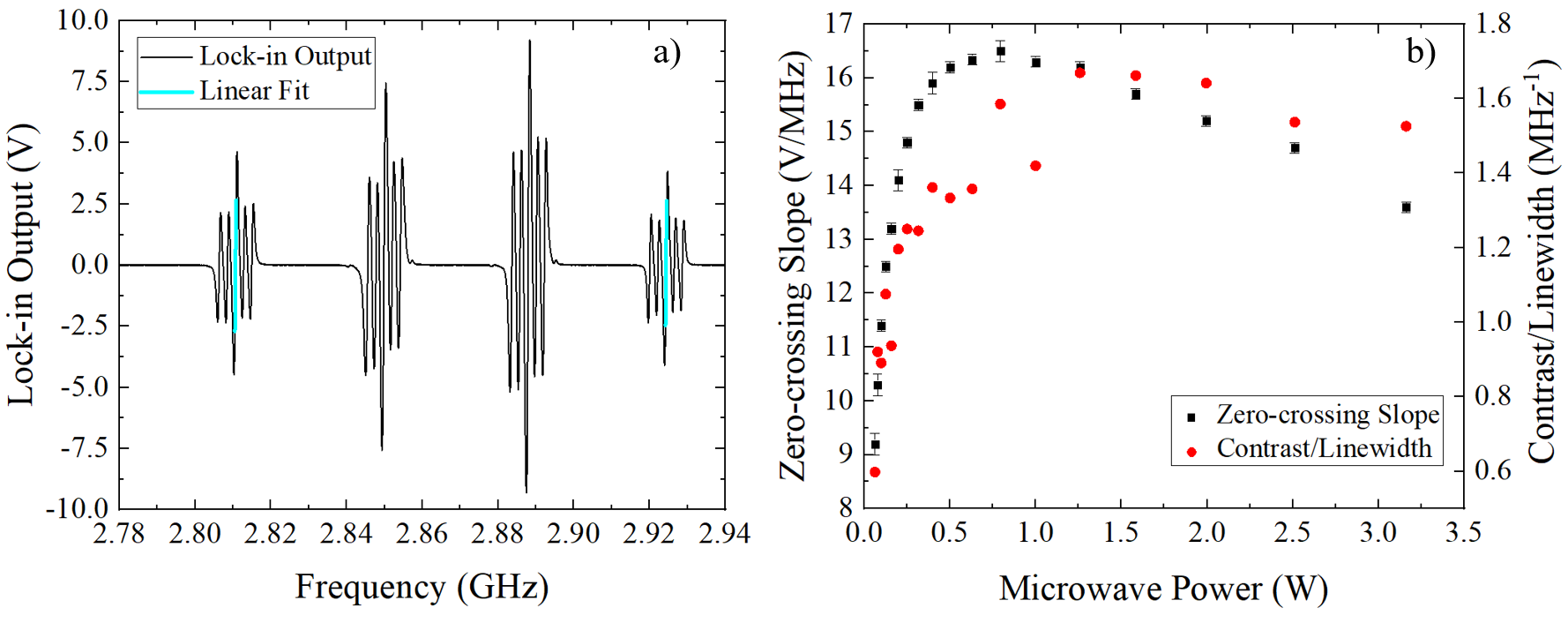}
\caption{a) An ODMR spectrum of the NVC as a function of varying microwave frequency. b) The zero-crossing slope and the contrast-to-linewidth ratio as a function of applied microwave power after amplification and neglecting cable losses. All measurements were taken using a frequency modulation depth of 300~kHz and a modulation frequency of 3.0307~kHz.}
\label{LIA_ODMR}
\end{figure}

Two methods were used to determine the sensitivity. The first involved linear fits to the outermost NVC lock-in derivative spectrum resonances to determine the calibration constant, see figure~\ref{LIA_ODMR}a) whilst the second involved applying known test fields to determine the magnetometer response. The maximum value of the zero-crossing slope, see figure~\ref{LIA_ODMR}b), was found to occur at microwave power of $\sim$0.8~W, a frequency depth of 300~kHz and a modulation frequency of 1.0307~kHz. Equation~\ref{PSNL} shows that the shot-noise limited sensitivity is improved by increasing the ratio of contrast/linewidth ($ C/\Delta\nu$), but there is a trade-off here as shown in figure \ref{LIA_ODMR}b) because increasing the microwave power can increase both the contrast and the linewidth. The linewidth and contrast were extracted directly from ODMR spectra prior to lock-in amplification when exciting all three nitrogen-14 hyperfine resonances. The trends inferred from the ($ C/\Delta\nu$) ratio in figure~\ref{LIA_ODMR}b) suggests further optimisation of the microwave parameters is possible to optimise the sensitivity. It was found that the zero-crossing slope increased with a decreasing modulation frequency (see appendix~\ref{Zero-crossing_Slope_vs_Modulation frequency}), however, beyond 3.0307~kHz the noise floor of the FFT also increased and thus the best sensitivity was found to be 171~pT$/\sqrt{\text{Hz}}$ at a modulation frequency of 3.0307~kHz. This value improves upon the 1.5~nT$/\sqrt{\text{Hz}}$ obtained with a fiber-coupled diamond magnetometer with the same technique for sensitivity measurement.

\begin{figure} 
\includegraphics[width=0.6\textwidth]{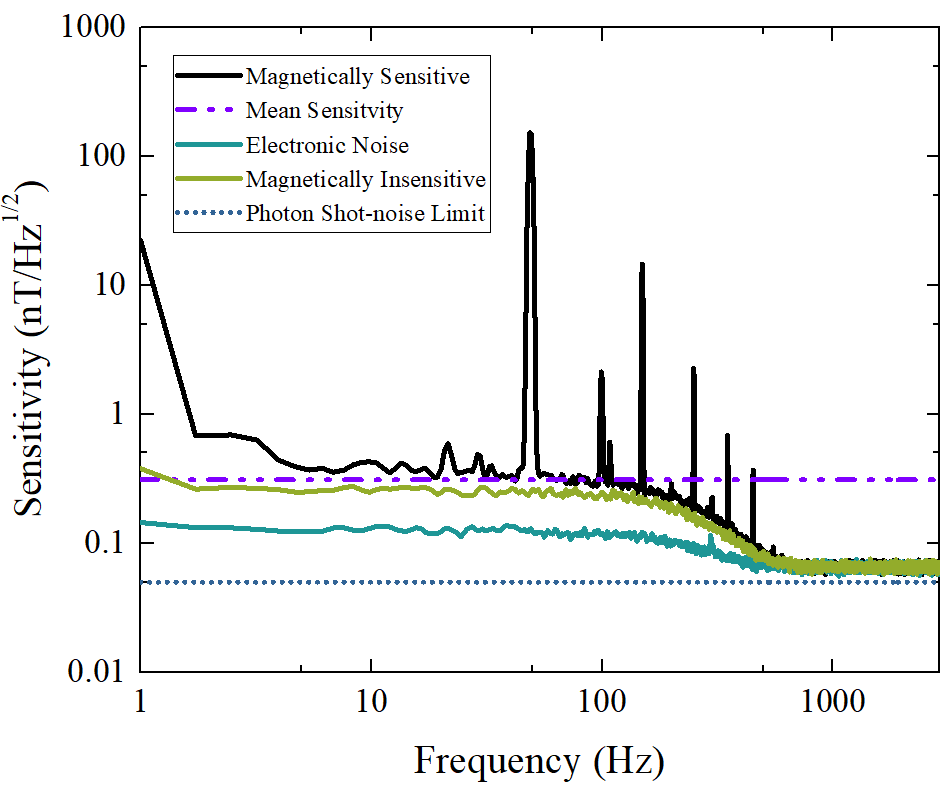}
\caption{Diamond magnetometry sensitivity spectrum: the mean sensitivity is 310~pT$/\sqrt{\text{Hz}}$ from 10-150 Hz. The noise floor is shown when no longer on a resonant frequency (magnetically insensitive) and with no applied laser or microwaves (electronic noise). The on resonance, off resonance and electronic noise are the mean of 160 1s FFT's.}
\label{Sensitivity_Image}
\end{figure}

For the second method known test fields were applied using a Helmholtz coil which was calibrated using a Hirst Magnetics GM07 Hall probe. The test fields were applied along (100) and the sensitivity was found to be (310~$\pm$~20)~pT$/\sqrt{\text{Hz}}$, as shown in figure~\ref{Sensitivity_Image}. The worse sensitivity is due to this non-optimal test field orientation. For the targeted application the fields of interest will be applied along the (100) direction and thus the sensitivity using the second method is considered to be the true sensitivity. Our sensitivity improves on the value of 35~nT$/\sqrt{\text{Hz}}$ previously obtained with a fiber coupled NVC magnetometer using applied test fields. The photon shot noise limit is calculated using equation~\ref{PSNL} from the fluorescence which was measured to be 1.2$\times 10^{15}$ photons/s by directly measuring the incident power level on a power meter (Thorlabs PM100D equipped with a power meter head Thorlabs S121C). The linewidth of 1.11~MHz and contrast of 1.76~\% were extracted directly from an ODMR spectrum prior to lock-in amplification when exciting all three nitrogen-14 hyperfine resonances. From this it is estimated that the photon shot noise limit is 50~pT$/\sqrt{\text{Hz}}$.

\section{Discussion}

Our sensitivity is 310~pT$/\sqrt{\text{Hz}}$ and thus we are a factor of $\sim$6 away from the shot-noise limit. This may be due to uncancelled laser and microwave noise some of which could be cancelled out through the implementation of a gradiometer which would also alleviate ambient magnetic noise from the environment~\cite{Blakley2015a,doi:10.1063/1.3507884}. To detect signals for MCG it is estimated that the sensitivity required would need to be over an order of magnitude beyond what we currently achieve~\cite{Clevenson2015,Watanabe2008}.

The biggest limitation of our system is the collection efficiency in which significant improvements are expected as the conversion efficiency of green to red photons is calculated to be 0.03\%. Improving this would also improve the excitation efficiency. Due to the high refractive index of diamond n$_{d}$~=~2.42, the majority of light emitted by the defects will undergo total internal reflection and thus the majority of emitted light will escape through the sides of the diamond~\cite{PhysRevB.85.121202}. A possible option for improvement would be an adaptation of the fluorescence waveguide excitation and collection~\cite{Zhang2018a} which reported a 96-fold improvement in the light collected. Another approach would be to surrounded the diamond with a total internal reflection lens to collect light from the diamond sides and focus it toward a small area~\cite{Xu2019}, which would be easier to integrate with our system, leading to an enhancement of 56 in the photon collection when compared to a lossless air objective of 0.55 N.A. This would represent a photon enhancement of $\sim$30 for our system and assuming a shot-noise limited scaling the measured sensitivity would become $\sim$60~pT$/\sqrt{\text{Hz}}$.

Ferrite flux concentrators have demonstrated a $\times$254 improvement in the sensitivity for a diamond magnetometer~\cite{Fescenko2019} at a cost of degrading the spatial resolution due to concentrating the flux from a large area and directing it toward a diamond. Due to the constraints of our system integrating the design discussed in~\cite{Fescenko2019} is not straightforward and thus the enhancement to sensitivity will be smaller. A further improvement would be to use the dual-resonance technique~\cite{Fescenko2019} which would  allow our system to be invariant to temperature fluctuations~\cite{PhysRevLett.104.070801} which is essential for practical applications of our magnetometer. 
Another way to introduce temperature invariance into our system would be the use of double-quantum magnetometry~\cite{Fang2013,Mamin2014a}. This would also be compatible with the use of pulsed schemes such as Ramsey magnetometry which would offer significant improvements to the sensitivity of a magnetometer compared to continuous wave excitation schemes~\cite{Barry14133,Barry2019}. However, it should be noted that significantly more laser excitation power and more homogeneous microwave driving fields will be required to realize the potential benefits of Ramsey magnetometry~\cite{Ahmadi2017,Jensen2014,Bauch2018}.

\section{Conclusion}

In this work a fiber-coupled magnetometer that reaches a sensitivity of (310$\pm$~20)~pT$/\sqrt{\text{Hz}}$ over the frequency range of 10-150~Hz has been presented. The mobility of the system and the compact nature of the sensor head are designed to target the application of magnetocardiography with further improvements discussed to be able to reach higher sensitivities.

\begin{acknowledgments}
The authors would like to thank Mareike Herrmann and Luke Johnson for materials processing, Jeanette Chattaway and Lance Fawcett of the Warwick Physics mechanical workshop, and Robert Day and David Greenshields of the Warwick Physics electronics workshop. We are grateful for insightful discussions with Matthew Turner, Danielle Braje, John Barry, Jennifer Schloss, Ronald Walsworth, Olga Young and Junichi Isoya. R. L. P. and G. A. S.'s PhD studentships are cofunded by the EPSRC Centre for Doctoral Training in Diamond Science and Technology (Grant No. EP/L015315/1). G. A. S.'s PhD studentship is additionally supported by Bruker. This work was supported by the EPSRC Quantum Technology Hub NQIT (Networked Quantum Information Technologies - Grant No. EP/M013243/1), QCS (Quantum Computing and Simulation - Grant No. EP/T001062/1) and funding from NICOP (Grant No. N62909-16-1-2111-P00002 - Towards a Picotesla DC Diamond Magnetometer). E. C. N and W. T. were supported by the Warwick University URSS (Undergraduate Research Support Scheme). B. L. G is supported by the Royal Academy of Engineering. G. W. M. is supported by the Royal Society.
\end{acknowledgments}

\appendix

\section{Diamond Characterisation}
\label{Diamond_characterisation}

\begin{figure}[b] 
\includegraphics[width=\textwidth]{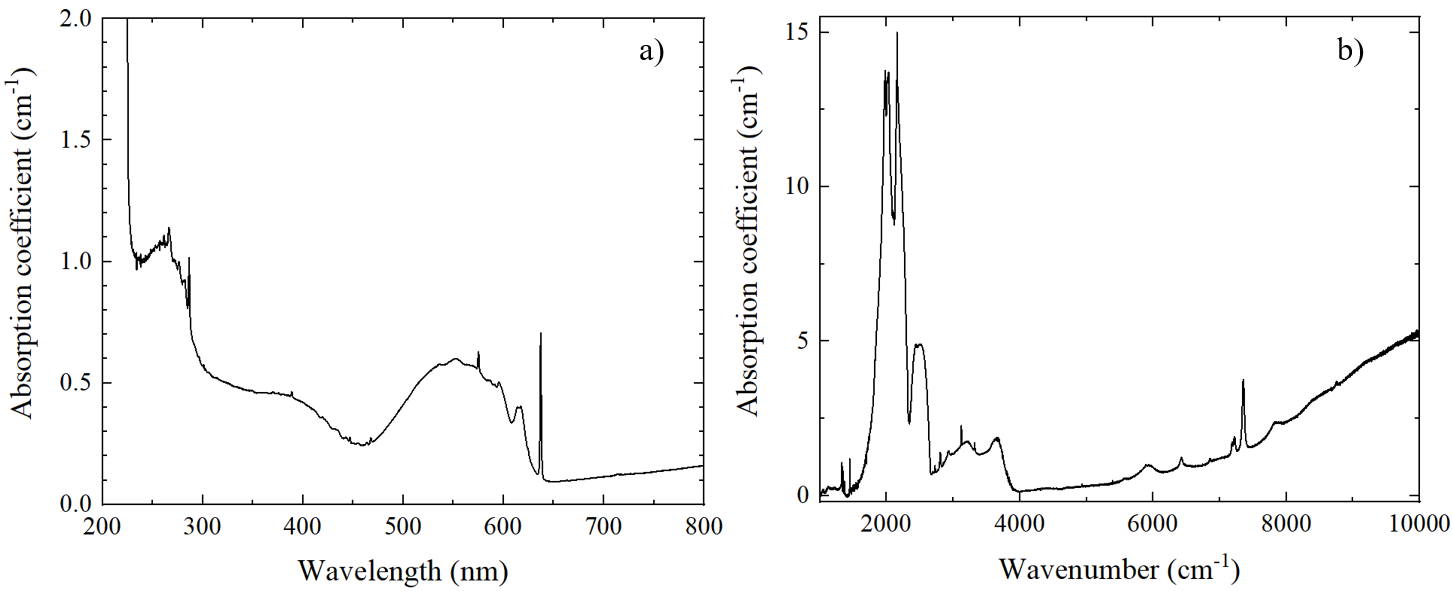}
\caption{a) The UV-Vis data to determine the concentrations of neutral and negatively charged NVC. b) FTIR data to determine the concentrations of neutral and positively charged substitutional nitrogen defects.}
\label{UV-VIS_and_FTIR}
\end{figure}

The defect concentration within the diamond was determined through Fourier-transform infrared spectroscopy (FTIR) and ultraviolet–visible spectroscopy (UV-Vis). US-Vis data, figure~\ref{UV-VIS_and_FTIR}a) were taken at 80~K on a Perkin Elmer Lambda 1050 Spectrometer equipped with an Oxford Instrument Optistat cryostat. The concentration was determined to be 4.6~ppm for negatively charged NVC and 0.8~ppm for neutral NVC and was found from the intensities of the 637~nm and 575~nm zero-phonon line respectively~\cite{Dale2015}. FTIR data, figure~\ref{UV-VIS_and_FTIR}b) were taken at room temperature using a Perkin Elmer Spectrum GX FT-IR spectrometer. The concentrations from FTIR were established to be 5.6~ppm for neutral substitutional nitrogen (N$_{s}^{0}$) and 3~ppm for positively charged substitutional nitrogen (N$_{s}^{+}$)~\cite{Lawson_1998,Liggins2010}. 

\section{Zero-crossing Slope vs. Modulation frequency}
\label{Zero-crossing_Slope_vs_Modulation frequency}

\begin{figure}[b] 
\includegraphics[width=0.5\textwidth]{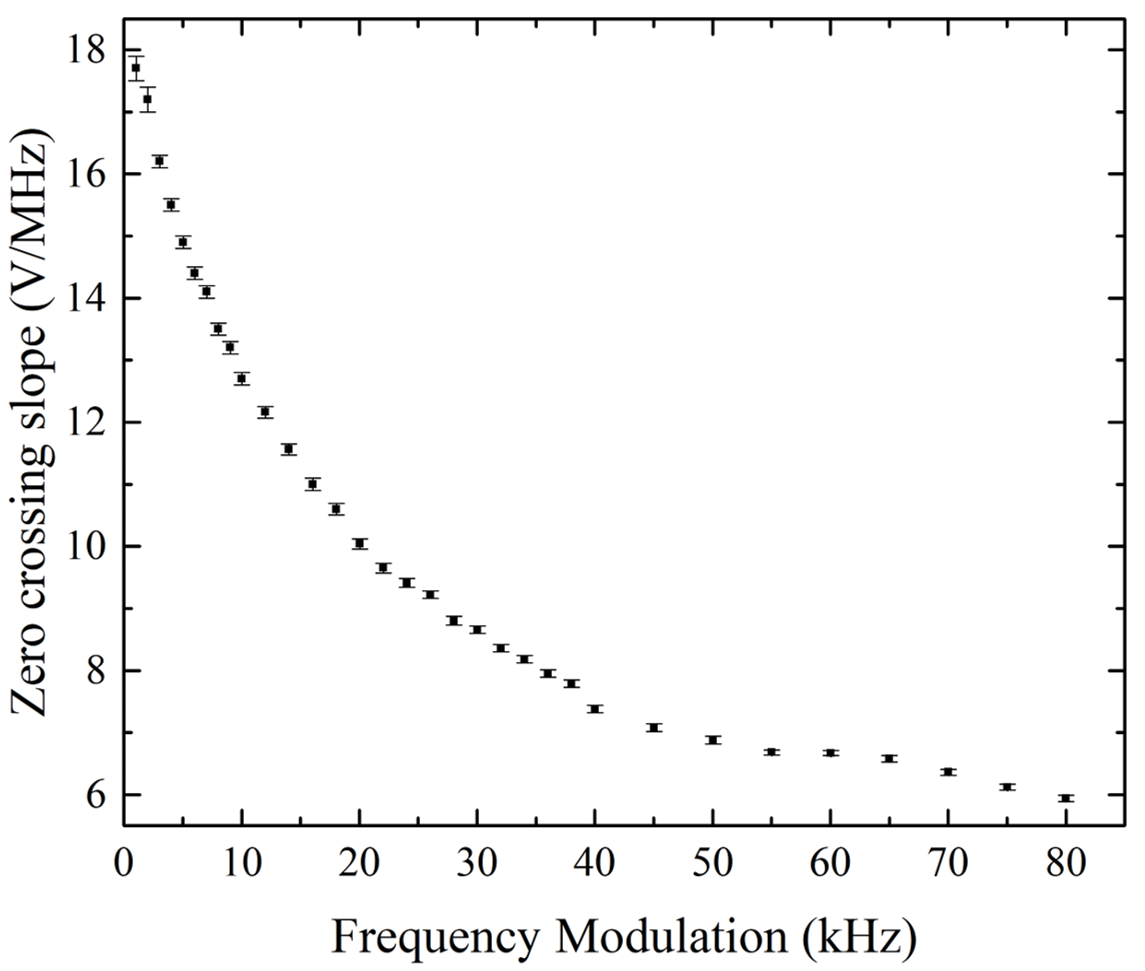}
\caption{The zero-crossing slope as a function of the modulation frequency.}
\label{MW_vs_ZCS}
\end{figure}

The variation of the zero-crossing slope as a function of the modulation frequency is shown in figure~\ref{MW_vs_ZCS}. The expected trend of a decrease in the zero-crossing slope for higher modulation frequencies due to the finite repolarisation time of the NVC centre is followed~\cite{Barry14133,El-Ella2017,PhysRevLett.106.030802}. Despite the continued increase of the zero-crossing slope at progressively lower modulation frequencies, the best sensitivity was achieved at a modulation frequency of 3.0307~kHz (data not shown), we attribute this to an increased susceptibility to noise at particularly low modulation frequencies nearer to DC. The maximum value of the zero-crossing slope at a modulation of 3.0307~kHz was 17.9~V/MHz which was slightly higher than the maximum in figure~\ref{LIA_ODMR}b) due to improved photon collection rate.

\bibliography{library}

\end{document}